# Extraordinarily Efficient Conduction in a Redox-Active Ionic Liquid

Verner K. Thorsmølle,*[a,b] Guido Rothenberger,[a] Daniel Topgaard,[c] Jan C. Brauer,[a] Dai-Bin Kuang,[d] Shaik M. Zakeeruddin,[a] Björn Lindman,[c] Michael Grätzel,[a] and Jacques-E. Moser[a]

[a] Dr. V. K. Thorsmølle, Dr. G. Rothenberger, J. C. Brauer, Prof. D.-B. Kuang, Dr. S. M. Zakeeruddin, Prof. M. Grätzel, Prof. J.-E. Moser
Laboratory for Photonics and Interfaces
Institute of Chemical Sciences and Engineering
École Polytechnique Fédérale de Lausanne
Station 6, CH-1015 Lausanne, (Switzerland)
Fax: (+41)216934111
E-mail: verner.thorsmolle@epfl.ch

[b] Dr. V. K. Thorsmølle
Département de Physique de la Matière Condensée
Université de Genève
24 quai Ernest-Ansermet, CH-1211 Genève, (Switzerland)

[c] Prof. D. Topgaard, Prof. B. Lindman
Department of Physical Chemistry 1
Lund University
SE-22100 Lund, (Sweden)

[d] Prof. D.-B. Kuang
School of Chemistry and Chemical Engineering
Sun Yat-Sen University
Guangzhou, Guangdong, 510275, (China)

**Iodine added to iodide-based ionic liquids leads to extraordinarily efficient charge transport, vastly exceeding that expected for such viscous systems. Using terahertz time-domain spectroscopy, in conjunction with dc conductivity, diffusivity and**

**viscosity measurements we unravel the conductivity pathways in 1-methyl-3-propylimidazolium iodide melts. This study presents evidence of the Grotthuss mechanism as a significant contributor to the conductivity, and provides new insights into ion pairing processes as well as the formation of polyiodides. The terahertz and transport results are reunited in a model providing a quantitative description of the conduction by physical diffusion and the Grotthuss bond-exchange process. These novel results are important for the fundamental understanding of conduction in molten salts and for applications where ionic liquids are used as charge-transporting media such as in batteries and dye-sensitized solar cells.**

## Introduction

The Grotthuss mechanism – a bond-exchange process – first introduced two centuries ago[1], has proved vital in explaining the high mobility of protons in water.[2] Similarly, this mechanism has been suggested to account for the high conductivity in molten polyiodides[3,4] with no defiant proof. Here we provide solid evidence that a Grotthuss bond-exchange mechanism is of fundamental importance for conduction in 1-methyl-3-propylimidazolium iodide (*PMII*) melts.

Ionic liquids are receiving increasing attention, owing to their unique properties such as high ionic conductivity, non-volatility and non-flammability, making them versatile alternatives to conventional solvent-based systems. Their potential applications range from electrolytes in solar cells[5], fuel cells[6], supercapacitors and batteries[7], lubricants and heat-transfer fluids[8] to solvents for clean chemical synthesis and catalysis.[9] Iodine addition to iodide-based ionic liquids leads to extraordinarily efficient charge transport, vastly exceeding that expected for such viscous systems.[10–15] Here we study *PMII* melts, whose structure is shown in Figure 1a. *PMII* is the benchmark of the iodide salts that form room temperature ionic liquids, characterized by a relatively low viscosity, and has therefore been the candidate of choice for binary ionic liquids in the solvent-free dye-sensitized Grätzel solar cell.[5]

The composition and the physical properties of the melts formed by the addition of iodine ($I_2$) to *PMII* depend on several chemical equilibria. *PMII* consists of a *PMI*$^+$ cation and an $I^-$ anion. Ion pairing (*PMI*$^+$ + $I^-$ ⇌ *PMII*) has an impact on the conductivity since *PMII* is uncharged. Triiodide ($I_3^-$) is formed by addition of iodine to the iodide anion $I^-$ ($I^-$ + $I_2$ ⇌ $I_3^-$). At higher $I_2$ concentrations, higher polyiodides, $I_5^-$, $I_7^-$, etc. appear via $I_3^-$ + $I_2$ ⇌ $I_5^-$, $I_5^-$ + $I_2$ ⇌ $I_7^-$, etc.[14] The bond exchange reaction $I^-$ + $I_3^-$ ⇌ $I_3^-$ + $I^-$ provides a Grotthuss conductivity mechanism in addition to the physical diffusion of ions, through which iodide and triiodide ions are displaced without effective mass transfer (see Figure 1b).

**Results and Discussion**

We characterized the dynamical behaviour of *PMII/$I_2$* mixtures using terahertz time-domain spectroscopy (THz-TDS).[16] The THz electric field was coherently measured after transmission through a 100-μm quartz cuvette containing the sample, the empty cuvette serving as reference. The time-domain waveforms were temporally windowed to eliminate the effects of multiple reflections within the cuvette walls, but not in the sample. The Fourier transformed waveform of the sample was then divided by that of the reference yielding the complex transmission $\tilde{t}(\omega)$ as a function of frequency, $\upsilon = \omega/2\pi$. The frequency-dependent complex dielectric function, $\tilde{\varepsilon}(\omega) = \varepsilon_1(\omega) + i\varepsilon_2(\omega)$, can then be extracted from $\tilde{t}(\omega)$. The complex conductivity, $\tilde{\sigma}(\omega) = \sigma_1(\omega) + i\sigma_2(\omega)$, follows from the relation $\tilde{n}^2 = \tilde{\varepsilon} = \varepsilon_\infty + i\tilde{\sigma}/\varepsilon_0\omega$, where $\tilde{n}$ is the complex refractive index, $\varepsilon_0$ the permittivity of vacuum and $\varepsilon_\infty$ the high-frequency dielectric constant.

Figure 2a shows the real part of the complex conductivity $\text{Re}(\tilde{\sigma}(\nu))$ at various $[I_2]_a$ concentrations measured at room temperature. The resonance feature centered at 2.03 THz for pure *PMII* decreases steadily in amplitude with increasing $[I_2]_a$ concentration until 4.2 M, corresponding to a nearly equimolar amount of *PMII* and iodine ($x_{I_2} \approx 0.5$). Upon further increasing $[I_2]_a$, the resonance maximum then shifts to 2.29 THz and increases in amplitude. Simultaneously, a noticeable increase in $\text{Re}(\tilde{\sigma}(\nu))$ is observed below 0.4 THz. From an analysis of the complex dielectric function for $x_{I_2} < 0.5$ (Supporting Information section S3), together with complementary studies of the dc conductivity (Figure 2b) shown below, it becomes clear that the 2.03 THz resonance results from the absorption of two species; namely the associated or ion-paired (IP) form of *PMII*, and the dissociated (DIS) forms *PMI$^+$/I$^-$* or *PMI$^+$/I$_3^-$*. $\tilde{\varepsilon}(\nu)$ for the various concentrations $[I_2]_a$ of iodine depends on the relative volume fractions and the dielectric functions $\tilde{\varepsilon}_{IP}(\nu)$ and $\tilde{\varepsilon}_{DIS}(\nu)$ of the two components IP and DIS of the mixture. (These dielectric functions do not depend on the iodine concentration and no distinction is made between the anions $I^-$ and $I_3^-$ as the imidazolium cation is assumed to be the main contributor to the signal[17]). The resonance at 2.29 THz indicates the formation of higher polyiodides above 4.2 M, which is supported by the coincidence of the maximum resonance frequency with that of the linear inner stretch of $I_5^-$ at ~2.25 THz.[18] Absorption due to higher polyiodides occurs at yet higher frequencies.[18,19]

The dc conductivity σ of the *PMII*/$I_2$ mixtures versus $[I_2]_a$ at room temperature is shown in Figure 2b. Starting at 0.27 mS/cm, it increases monotonically until $[I_2]_a \sim 4.4$ M, coinciding with $x_{I_2} \approx 0.5$. Thereafter it increases abruptly reaching a value of 21.5 mS/cm at $[I_2]_a \sim 5.6$ M, which is almost two orders of magnitude higher compared to the iodine-free melt. The steady increase of σ in the $I_2$ concentration range where $x_{I_2} < 0.5$ coincides with the steady decrease of the 2.03 THz resonance (Figure 2a), whereas the appearance and increase of the 2.29 THz resonance, indicative of higher polyiodides, coincides with the sharp increase in the dc conductivity for $x_{I_2} > 0.5$. For $x_{I_2} < 0.5$, σ is determined by the $I^-/I_3^-$-ratio as counterions to *PMI*$^+$, whereas for $x_{I_2} > 0.5$ higher polyiodides appear as counterions. In order to rationalize the observed conductivity behaviour in the lower range of iodine concentrations, up to $[I_2]_a = 3.6$ M, where effects due to the formation of the higher polyiodides beyond $I_3^-$ can be neglected, we establish a conductivity model where the species present in the melt are restricted to *PMII*, *PMI*$^+$, $I_2$, $I^-$ and $I_3^-$. This model, outlined in Supporting Information section S2.1, considers physical diffusion as well as the bond exchange mechanism. The dc conductivity of the *PMII*/$I_2$ mixture is determined by the interplay of several factors: (i) a dilution effect when adding iodine; (ii) the viscosity dependence on $[I_2]_a$ of the solution; (iii) the equilibrium between iodine, iodide and triiodide; (iv) ion pairing; (v) the bond exchange reaction. The conductivity is then expressed as

$$\sigma = \frac{e_o^2 N_{AV}}{6} \left( \frac{1}{\pi \, \eta([I_2]_a)} \left( \frac{c_{PMI^+}}{r_{PMI^+}} + \frac{c_{I^-}}{r_{I^-}} + \frac{c_{I_3^-}}{r_{I_3^-}} \right) + \frac{k_{ex} \delta^2}{k_B T} c_{I^-} c_{I_3^-} \right). \qquad (1)$$

Here, $e_o$ is the elementary electron charge, $N_{AV}$ is Avogadro's number, $k_B$ is the Boltzman constant, $\eta([I_2]_a)$ is the concentration-dependent viscosity, and $c$ and $r$ are the concentration and hydrodynamic radius of an ion, $k_{ex}$ is the second-order rate constant of the iodide-triiodide bond exchange reaction, $\delta^2 = \delta_{I^-}^2 + \delta_{I_3^-}^2$, and $T$ is the temperature. The first term in the parenthesis of equation (1) describes the physical diffusion of $PMI^+$, $I^-$ and $I_3^-$ ions, and the last term is the Grotthuss bond exchange contribution. We assume that the rate constant $k_{ex}$ for the bond exchange mechanism between $I^-$ and $I_3^-$ equals Smoluchowski's expression for a diffusion controlled reaction times a constant $f$ (see Supporting Information section S2.1).

Next, we perform a least squares fit of equation (1) to the dc conductivity data (Figure 2b) up to 3.6 M, where the two parameters to be optimized are the ion pairing equilibrium constant, $K_1$ of the reaction $PMII \rightleftharpoons PMI^+ + I^-$, which determines the concentrations of the ions and $f\delta^2$ which determines the efficiency of the exchange mechanism. A least squares minimum is found at $K_1$ = 5.2 M and $f\delta^2$ = 7.2×10$^{-19}$ m$^2$. Knowing $K_1$, one can calculate the concentrations of the different species in the solution (Figure 3a). With an estimated value of $\delta$ = 9.3 Å, the factor $f$ is 0.83, from which we conclude that the exchange reaction is close to diffusion controlled, which implies that the time scale for the bond exchange reaction is in the ns range ($k_{ex} = 1 \times 10^8 \; M^{-1} \cdot s^{-1}$ at $[I_2]_a$ = 3.6 M). The calculated dc conductivity is shown in Figure 3b. The excellent agreement between the calculated and the measured dc conductivity lends support to the validity of the proposed model, which excludes an electronic contribution. In particular, the Grotthuss exchange mechanism emerges as a key factor in enhancing the conductivity of the melts. According to the present model, the Grotthuss effect increases the conductivity of $PMII$ at 3 M iodine concentration by as much as 50%. Above ~3.6 M exchange reactions between higher polyiodides may become important, which further enhance the Grotthuss contribution to ionic transport.[20] A hysteresis loop in the temperature dependence supports the exclusion of electronic transport (See Supporting Information, Figure S2,).

The sharp increase in σ for $x_{I_2} > 0.5$ cannot be explained by simple ionic conduction alone. In this regime, the larger and less mobile higher polyiodides, $I_5^-$, $I_7^-$, etc. become increasingly dominant among the iodine species, at the expense of the smaller and more mobile iodide and triiodide ions $I^-$ and $I_3^-$. This, together with the dilution effect would tend to decrease σ in terms of physical diffusion when iodine is added to the melt. The viscosity decreases slightly in this concentration range (see Supporting Information, Figure S1) which would cause a small gradual increase of σ. Thus, these factors cannot cause the observed sudden increase of σ for $x_{I_2} > 0.5$, the origin of which we attribute to enhanced Grotthuss bond exchange among the higher polyiodides. It is highly likely that the increased iodine/iodide packing density enhances the Grotthuss mechanism due to the reduced distance between the iodide/polyiodide species involved in the bond exchange. The threshold concentration of $x_{I_2} \sim 0.5$ corresponds to 48 wt % or 27 % volume fraction (59 wt % or 37 % volume fraction at 5.6 M), implying a considerably densely-packed polyiodide medium which may support Grotthuss bond exchange among the higher polyiodides in addition to only between $I^-$ and $I_3^-$.

Figure 2c shows the room temperature diffusivity of the $PMI^+$ species in $PMII/I_2$ mixtures as a function of $[I_2]_a$ measured using pulsed gradient spin echo $^1$H NMR. The experimentally observed $PMI^+$ diffusivity is a population weighted average of the contributions from the $PMI^+$ cation and the ion-paired form $PMII$ on account of the fast chemical exchange between the various species on the 20 ms time scale defined by the NMR experiment. The diffusivity increases nearly exponentially, with an inflection of the slope near 4.2 M. The increase in diffusivity is partly due to the decrease in viscosity. In addition, it is affected by the ion pairing equilibrium, $PMII \rightleftharpoons PMI^+ + I^-$. With increasing $[I_2]_a$ the bulkier $PMII$ is replaced by the more mobile $PMI^+$ ion, up to 4.2 M, where the ion-paired form of *PMII* vanishes (Figure 3a). The slight raise of the diffusivity above 4.2 M is then only due to a reduction of the viscosity in a melt that contains only $PMI^+$ as cation.

Having determined $K_1$, the equilibrium constant for ion pairing, one can calculate the concentrations of the IP and DIS forms of *PMII* that are needed for the analysis of the THz data. (See Supporting Information section S3). The decrease of the resonance amplitude at 2.03 THz resonance (Figure 2a) with increasing $[I_2]_a$, is correlated with the decrease of the concentration of the associated or ion-paired (IP) form of *PMII* and the increase of the dissociated (DIS) forms of $PMI^+/I^-$ or $PMI^+/I_3^-$ (Figure 3a). From the complex dielectric function of these THz spectra one can extract the separate dielectric functions for both the IP and the DIS forms (see Supporting Information, Figure S6). These calculated dielectric functions can each be modeled by two relaxation processes: relaxation of the orientational polarization induced by the electromagnetic radiation and resonance-induced polarization (plasma relaxation due to freely moving electrons is not considered). The former is described with a Debye relaxation time of 1.21 ps (0.99 ps) for IP (DIS), while the latter is described by classical damped harmonic oscillators which may be due to inter-ion vibrations. The successful analysis of the THz data, with the correlation between the 2.03 THz resonance and the IP/DIS forms of *PMII*, substantiates the notion of ion pairing which is one of the cornerstones in the proposed model.

**Conclusion**

In conclusion, we have provided a quantitative description of the conductivity pathways in the *PMII* ionic liquid as a function of iodine concentration, which highlights the importance of ion pairing and the Grotthuss bond exchange mechanism in addition to physical diffusion. Our model is valid for concentrations up to ~3.6 M, explaining both the dc conductivity, the diffusivity and the THz data. Results obtained at concentrations >3.6 M support the appearance of higher polyiodides giving rise to increased bond-exchange. The sudden increase in the dc conductivity at a threshold concentration of $[I_2]_a$ ~ 4.2 M coincides exactly with the displacement of the resonance feature from 2.03 THz to 2.29 THz in the THz data. The present results provide a self-consistent picture of the fundamental mechanisms responsible for the conductivity in the *PMII* melts that firmly places the Grotthuss bond exchange mechanism as an important conduction pathway. It enhances the conductivity by 50% already at 3 M, becoming the dominant contribution to the conductivity at 4.2 M. These novel results are important for the fundamental understanding of conduction in molten salts and for applications where ionic liquids are used as charge-transporting media such as in batteries and dye-sensitized solar cells.

## Experimental Section

**Preparation of *PMII* samples.** The ionic liquid, 1-propyl-3-methylimidazolium iodide (*PMII*), was prepared according to the procedure reported in Ref. [21]. The *PMII*/$I_2$ mixtures were prepared by adding the calculated amount of iodine to the *PMII*, and the solution was then stirred with a magnetic stirring palette for two days to ensure that the iodine is completely dissolved.

**dc conductivity measurements.** A Radiometer Analytical CDM210 conductivity meter was used for the dc conductivity measurements. A Radiometer CDC749 conductivity cell with a nominal cell constant of 1.70 cm$^{-1}$ was calibrated with a 0.1 M KCl aqueous solution in preparation for the experiments. About 0.2 ml of the *PMII*/$I_2$ mixture was introduced into a glass tube and solvated gases were removed using a slight vacuum of 0.6 Torr at room temperature for about 30 minutes. The conductivity cell and the *PMII/I$_2$* mixture were then inserted into a glass tube in an Oxford cryostat with liquid nitrogen cooling for temperature control. At each temperature the *PMII/I$_2$* mixture was allowed to equilibrate until the dc conductivity reading had stabilized, which would usually take between 30 minutes and 1 hour. The dc conductivity values were controlled by impedance spectroscopy using an Autolab Frequency Analyzer setup, which consists of an Autolab PGSTAT 30 and a frequency-response analyzer module combined with a thermostatic chamber capable of reaching -40 ºC.

**Viscosity measurements.** The viscosity measurements were carried out using a Rheometrics ARES Rheometer. All measurements were performed using parallel plates with a diameter of 25 mm, a gap of 0.2 mm, and a frequency of 1 Hz.

**Differential scanning calorimetry measurements.** The differential scanning calorimetry (DSC) measurements were carried out using a TA Q100 apparatus.

**Terahertz time-domain spectroscopy measurements.** The terahertz time-domain spectroscopy (THz-TDS) measurements were carried out with an electro-optical terahertz setup which is described in Ref. [22] and in a movie on the website: http://videoserv.epfl.ch:8080/ramgen/Archive/Nas/Thz_spectro.rm. The THz-TDS experiment utilizes a CPA-2001 amplified Ti:Sapphire laser system from Clark-MXR providing 120 fs pulses at a repetition rate of 1 kHz and at a wavelength of 778 nm.

**Nuclear magnetic resonance diffusivity measurements.** NMR experiments were performed on a Bruker Avance-II spectrometer operating at a $^1$H resonance frequency of 200.13 MHz. Pulsed field gradients were generated by a Bruker DIF-25 probe driven by a GREAT 40 amplifier. The sample temperature was controlled to an accuracy of 0.5ºC. The self-diffusion of *PMI*$^+$ was monitored by following the $^1$H NMR signal in a pulsed gradient spin echo experiment[23-25] using a gradient pulse length $\delta$ = 4.2 ms, time between onset of gradient pulses $\Delta$ = 24.4 ms, and gradient strength *G* incremented from 1 to 100% of the maximum value 9.6 T/m in 16 logarithmically spaced steps. The self-diffusion coefficient *D* was evaluated by fitting

$$I = I_0 \exp\left[-\gamma^2 G^2 \delta^2 \left(\Delta - \delta/3\right) D\right] \qquad (2)$$

to the experimental signal intensities *I*. In equation (2), $I_0$ is the signal intensity at zero gradient strength and $\gamma$ is the magnetogyric ratio (2.675·10$^8$ radT$^{-1}$s$^{-1}$ for $^1$H). Monoexponential signal decay was observed for all concentrations and temperatures indicating chemical exchange between the IP and DIS forms of *PMII* on a time scale much shorter than the 20 ms observational time scale of the NMR experiment.

## *Acknowledgements*

*We thank Paul Dyson, Péter Péchy, Andrzej Sienkiewicz, Véronique Michaud and Christoper Plummer for valuable discussions or technical assistance. Work at EPFL was supported by the Swiss National Science Foundation and the Marie Curie Reintegration Grant MIRG-CT-2005-014868. Work at LU was supported by the Swedish Research Council through the Linnaeus Center of Excellence on Organizing Molecular Matter.*

**Keywords:** ionic liquids · Grotthuss conduction · ion pairs · terahertz spectroscopy · electrolyte

**Figure Legends**

Figure 1. *PMII* molecular structure and illustration of Grotthuss bond exchange mechanism. (a) Molecular structure of 1-methyl-3-propyl-imididazolium iodide (*PMII*). (b) Illustration of Grotthuss bond exchange mechanism between an iodide and a triiodide ion. The iodide (blue) in close proximity to the one end of a triodide chain (red) forms a loose bond (dashed line). The exchange reaction results in the strengthening of this bond while an iodide is liberated at the other end. This leads to the displacement of iodide by $\delta_{I^-}$ and of triodide by $\delta_{I_3^-}$ as indicated. See also Supporting Information section S2.1.

Figure 2. THz, dc conductivity and diffusivity measurements of *PMII*/$I_2$ mixtures. (a) Terahertz data showing the real part of the complex conductivity as a function of frequency at various iodine concentrations $[I_2]_a$ (23.7 °C). The upper inset shows the evolution of the maximum value of the resonance at 2.03 THz (blue dots) and 2.29 THz (red dots). The dotted black line is calculated from the conductivity model (see text). (b) Measured dc conductivity versus $[I_2]_a$ (23 °C). (c) Diffusivity of *PMI*$^+$ (red dots) versus iodine concentrations $[I_2]_a$ measured by NMR (20.5 °C). The blue lines are fits to an exponential $D = D_0 \exp\left([I_2]_a / C_0\right)$ with $C_0$ = 1.54 M ($[I_2]_a$ < 4.2 M) and $C_0$ = 4.20 M ($[I_2]_a$ > 4.2 M.

Figure 3. dc conductivity compared to model. (a) Concentrations of the ion-paired salt, *PMII*, of iodine $I_2$ and of the dissociated ions *PMI*$^+$, $I^-$, and $I_3^-$ calculated as a function of analytical iodine concentration $[I_2]_a$, using $K_1$ = 5.2 M and $K_2$ = 1.0×10$^8$ M$^{-1}$ (see Supporting Information section S2.2). (b) Measured dc conductivity (red circles) versus $[I_2]_a$ and calculated dc conductivity (blue line) according to the proposed model

(equation (1)). The dashed blue line shows the contribution to σ coming from the Grotthuss bond exchange mechanism.

**Figures**

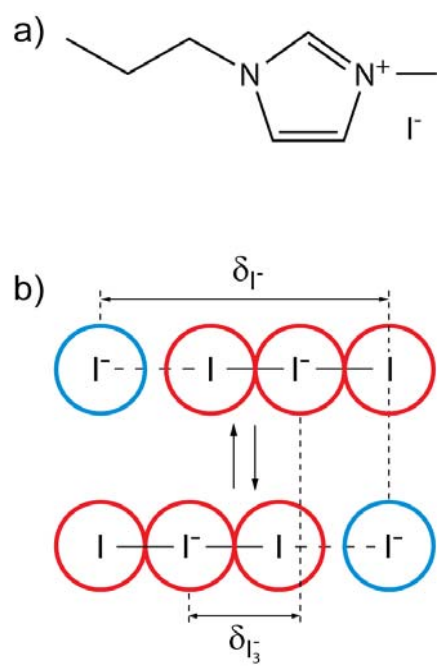

**Figure 1.**

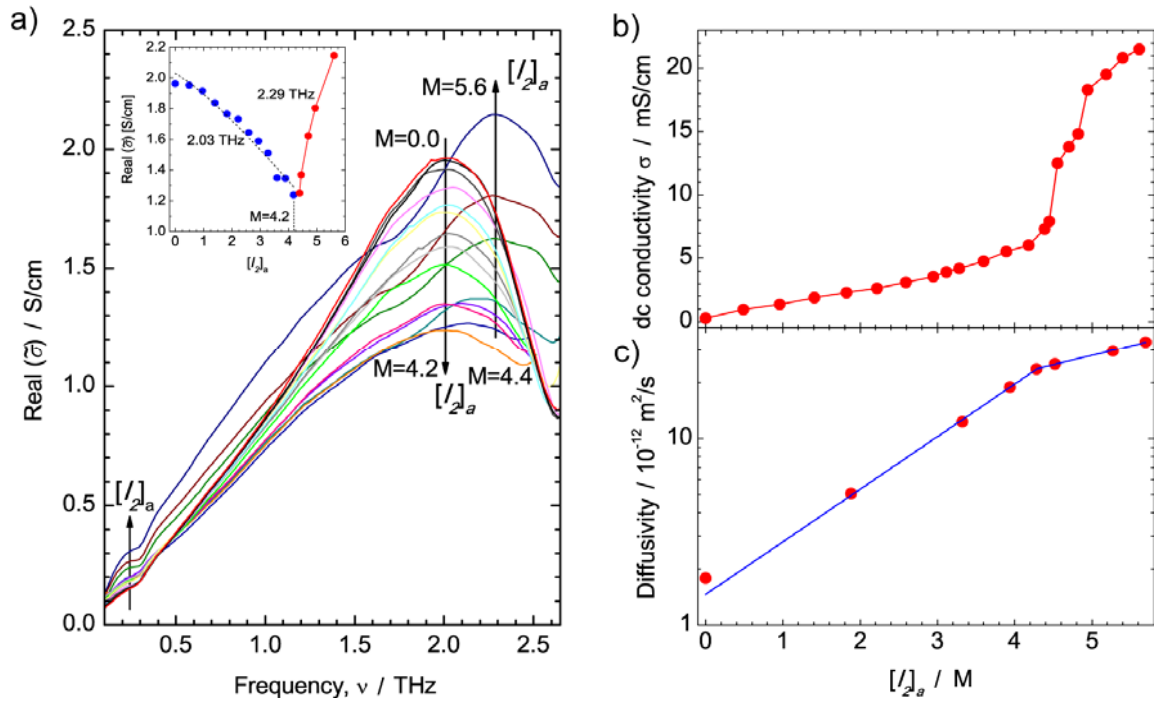

**Figure 2.**

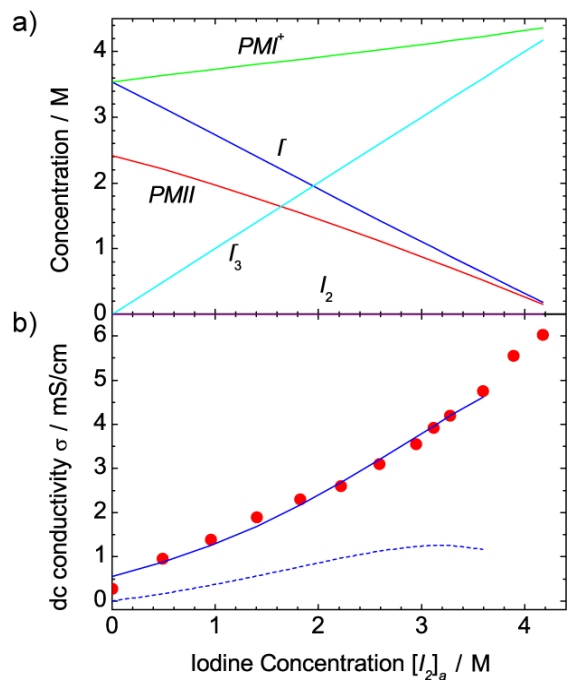

**Figure 3.**

**Table of Contents**

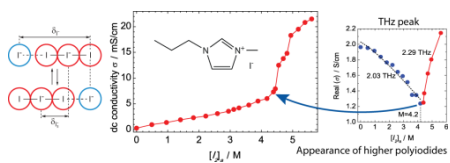

Extraordinarily Efficient Conduction in a Redox-Active Ionic Liquid

Grotthuss conduction and Ion-pairing in *PMII* melts versus iodine concentration. A sudden increase in the dc conductivity at a threshold concentration of $[I_2]_a \sim 4.2$ M coincides exactly with the displacement of the resonance feature from 2.03 THz to 2.29 THz in the THz data; tracking the decrease of the ion-paired form of *PMII* and the appearance of higher polyiodides which gives rise to increased bond-exchange.

Verner K. Thorsmølle,*[a,b] Guido Rothenberger,[a] Daniel Topgaard,[c] Jan C. Brauer,[a] Dai-Bin Kuang,[d] Shaik M. Zakeeruddin,[a] Björn Lindman,[c] Michael Grätzel,[a] and Jacques-E. Moser[a]

# Supporting Information

# Extraordinarily Efficient Conduction in a Redox-Active Ionic Liquid

Verner K. Thorsmølle*[a,b], Guido Rothenberger[a], Daniel Topgaard[c], Jan C. Brauer[a], Dai-Bin Kuang[a,d], Shaik M. Zakeeruddin[a], Björn Lindman[c], Michael Grätzel[a], Jacques-E. Moser[a]

[a]*École Polytechnique Fédérale de Lausanne, Institute of Chemical Sciences and Engineering, CH-1015 Lausanne, Switzerland.*

[b]*Département de Physique de la Matière Condensée, Université de Genève, 24 quai Ernest-Ansermet, CH-1211 Genève, Switzerland.*

[c]*Lund University, Department of Physical Chemistry 1, SE-22100 Lund, Sweden.*

[d]*School of Chemistry and Chemical Engineering, Sun Yat-Sen University, Guangzhou, Guangdong, 510275, China.*

*e-mail: verner.thorsmolle@epfl.ch

## Contents

## Section S1. Viscosity and dc conductivity measurement

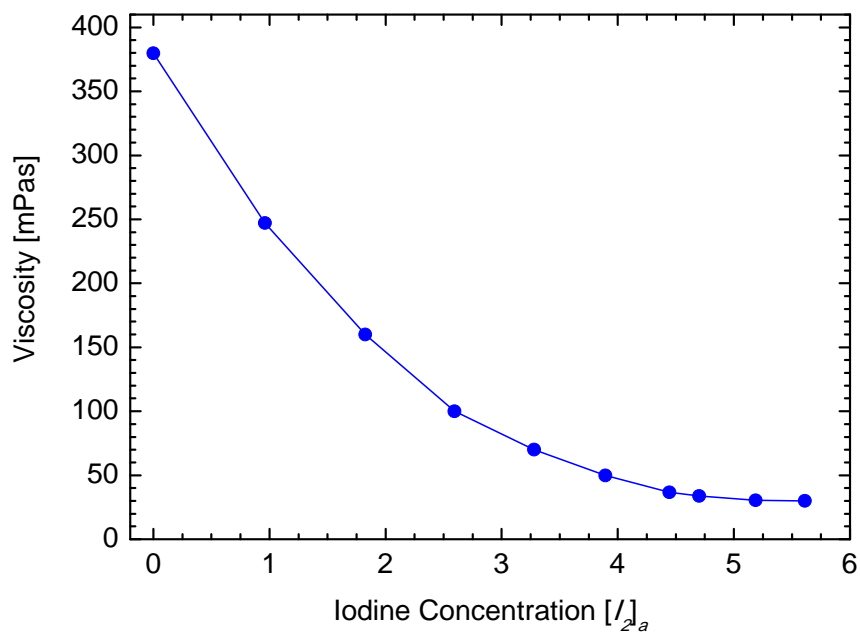

**Figure S1**. Measured viscosity versus iodine concentration $[I_2]_a$ at room temperature (23.7 °C).

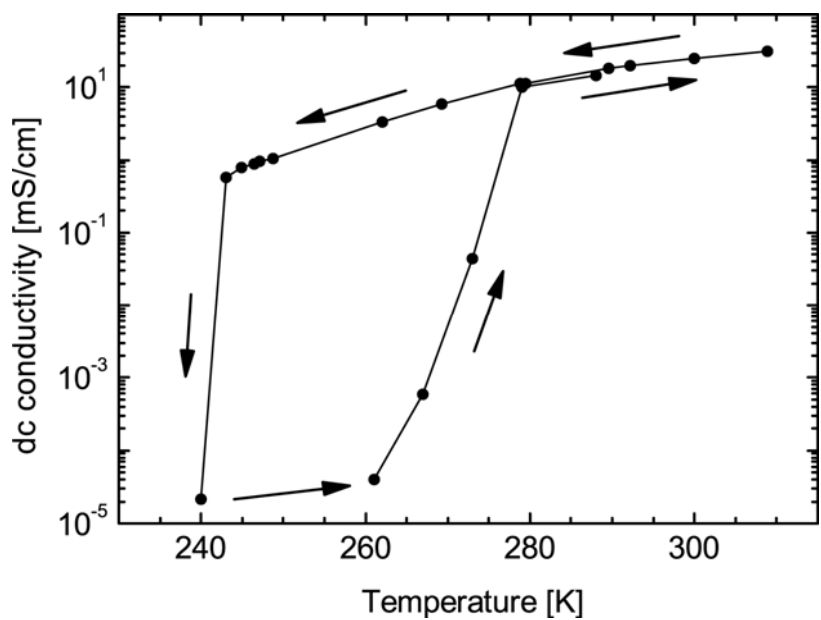

**Figure S2**. Temperature-dependent measurement of the dc conductivity for $[I_2]_a$ = 5.61 M.

The viscosity of *PMII* as a function of iodine concentration at room temperature is shown in Figure S1. The temperature dependence of the dc conductivity reveals a hysteresis loop for $[I_2]_a > 3.9$ M, which is shown for $[I_2]_a = 5.61$ M in Figure S2.

## Section S2. dc conductivity model

### S2.1 Derivation

Adding iodine to *PMII* melts affects the conductivity of the solution in several ways:

(i) Adding iodine does not provide additional charged species, but the volume increase diminishes their concentration. The calculation of the analytical iodine concentration $[I_2]_a$ takes this volume increase into account as the density of the solutions is known.

(ii) The viscosity decreases nearly exponentially with iodine concentration (Figure S1). This is attributed to the larger radius of the triiodide ions compared to iodide thus reducing the electrostatic attraction to the *PMI*$^+$ cations and hence the ion pairing. This effect results in an increased conductivity.

(iii) Given that the equilibrium $I^- + I_2 \rightleftarrows I_3^-$ is displaced towards the triiodide ion, adding iodine leads to an increase of the concentration of the larger triiodide ions at the expense of the smaller iodide ions. This effect would tend to decrease the conductivity, thus partly counterbalancing the viscosity effect (ii).

(iv) In solvent-free ionic liquids, cations and anions are in close proximity, rendering ion pairing highly probable. Ion pairing between the *PMI*$^+$ cation and the $I^-$ anion, *PMI*$^+$ + $I^- \rightleftarrows$ *PMII* tends to decrease the conductivity since the associated *PMII* ion-pair has no overall charge.

(v) The presence of both iodide and triiodide facilitates a Grotthuss-like bond exchange mechanism via the reaction

$$I^- + I_3^- \rightleftarrows I_3^- + I^-, \tag{S1}$$

which has previously been used to qualitatively explain the anomalous transport behaviour in $I^-/I_3^-$ containing systems[1-5]. The triiodide (iodide) approaches iodide (triiodide) from one end forming an encounter complex, from which triiodide (iodide) is released at the other end, without the ion having to cross that distance. The bond exchange reaction can be envisioned as a random walk, where the hopping of an "$I_2$" molecule from a triiodide to an iodide site occurs. This random walk can be described by an exchange diffusion constant $D_{ex,i}$. As illustrated in Figure 1b, a bond exchange event moves the iodide ion by a distance $\delta_{I^-}$ which is roughly equal to three I-I bond lengths in the triiodide ion, whereas the triiodide ion is displaced by a distance $\delta_{I_3^-}$ that is nearly equal to one I-I bond length. The enhancement of the iodide and triiodide diffusion tends to increase the overall conductivity.

The concentration of polyiodides higher than $I_3^-$ is negligible for $[I_2]_a < 3.6$ M. Here, we establish a conductivity model restricting the species to $PMII$, $PMI^+$, $I_2$, $I^-$ and $I_3^-$. Taking factors (i)-(v) into account, this model considers physical diffusion as well as the bond exchange mechanism. The ionic conductivity is given by the Nernst-Einstein equation

$$\sigma = \frac{F^2}{RT} \sum_i z_i^2 D_i c_i, \tag{S2}$$

where $F$ is the Faraday constant, $R$ the molar gas constant, $T$ the temperature, $D_i$ and $c_i$ the diffusion constant and concentration of ion $i$, respectively, and $z_i$ the absolute value of its charge. The physical diffusion of the ions is described by the Stokes-Einstein equation (stick boundary conditions)

$$D_{phys,i} = \frac{k_B T}{6\pi \eta([I_2]_a) r_i}. \qquad (S3)$$

Here, $k_B$ is the Boltzman constant, $r_i$ the hydrodynamic radius of an ion and $\eta([I_2]_a)$ is the concentration-dependent viscosity. The exchange diffusion constant can be defined as $D_{ex,i} = \delta_i^2/6\tau_i$ (for a 3-dimensional isotropic random walk), where $i$ now refers only to $I^-$ and $I_3^-$. Here, $\tau_i$ is the average time between two consecutive exchange events. $\tau_i$ depends on the rate of formation of the precursor complex, $d[complex]/dt = k_{ex}[I^-][I_3^-]$, where $k_{ex}$ is the rate constant of the iodide-triodide bond exchange. Assuming a pseudo-first order condition, one obtains

$$\frac{1}{\tau_{I^-}} = k_{ex}[I_3^-] \quad and \quad \frac{1}{\tau_{I_3^-}} = k_{ex}[I^-] \qquad (S4)$$

leading to $D_{ex,I^-} = \frac{k_{ex}\delta_{I^-}^2}{6}[I_3^-]$ and $D_{ex,I_3^-} = \frac{k_{ex}\delta_{I_3^-}^2}{6}[I^-]$. Since the physical diffusion can also be described by a random walk model, and assuming both the exchange and the physical diffusion processes take place simultaneously and independently, the effective mean square displacement can be expressed as $\langle r^2(t) \rangle = \langle r_{phys}^2(t) \rangle + \langle r_{ex}^2(t) \rangle$, leading to the Dahms-Ruff equation[6-9] for the effective diffusion constant, $D_i = D_{phys,i} + D_{ex,i}$. Taking all ions into account, the conductivity is expressed as

$$\sigma = \frac{F^2}{RT}\left[D_{phys,PMI^+} c_{PMI^+} + \left(D_{phys,I^-} + \frac{k_{ex}\delta_{I^-}^2}{6} c_{I_3^-}\right) c_{I^-} + \left(D_{phys,I_3^-} + \frac{k_{ex}\delta_{I_3^-}^2}{6} c_{I^-}\right) c_{I_3^-}\right], (S5)$$

or

$$\sigma = \frac{e_o^2 N_{AV}}{6} \left( \frac{1}{\pi \eta([I_2]_a)} \left( \frac{c_{PMI^+}}{r_{PMI^+}} + \frac{c_{I^-}}{r_{I^-}} + \frac{c_{I_3^-}}{r_{I_3^-}} \right) + \frac{k_{ex} \delta^2}{k_B T} c_{I^-} c_{I_3^-} \right). \quad (S6)$$

Here, $e_o$ is the elementary electron charge, $N_{AV}$ is Avogadro's number, and $\delta^2 = \delta_{I^-}^2 + \delta_{I_3^-}^2$. The first term in the outer parenthesis is the contribution of the physical diffusion of $PMI^+$, $I^-$ and $I_3^-$ ions to the conductivity, and the last term is the Grotthuss bond exchange contribution. This term shows that the diffusivity enhancement due to the bond exchange mechanism cannot be estimated separately for the $I^-$ and $I_3^-$ ions by measuring the conductivity of the solution. $\delta^2$ is about 10 times the square of one I-I bond length, ~2.93 Å[10], and thus $\delta$ = 9.3 Å. The hydrodynamic radii of the ions are $r_{PMI^+}$ = 3.49 Å [11], $r_{I^-}$ = 2.2 Å [12], and $r_{I_3^-}$ = 3.35 Å (estimated by the radius of gyration of a cylinder). We assume that the rate constant for the bond exchange mechanism $k_{ex}$ between $I^-$ and $I_3^-$ equals Smoluchowski's expression for a diffusion controlled reaction $k_{diffusion} = 4\pi N_{av}(D_{I^-} + D_{I_3^-})(r_{I^-} + r_{I_3^-})$ [13] times a constant $f$, leading to,

$$k_{ex} = f \times k_{diffusion} = \frac{2}{3} f \frac{RT}{\eta([I_2]_a)} \frac{(r_{I^-} + r_{I_3^-})^2}{r_{I^-} r_{I_3^-}}, \quad (S7)$$

where $R$ is the molar gas constant. The factor $f$ may include effects due to coulombic repulsion between ions of the same sign or the presence of a barrier of activation for the exchange reaction.

**S2.2 Fitting procedure**

In order to find the concentrations of the species *PMII*, $PMI^+$, $I_2$, $I^-$ and $I_3^-$ we consider the equilibria between the ion-paired and the dissociated forms of *PMII*, as well as between iodide and triiodide.

$$PMII \underset{k_{-1}}{\overset{k_1}{\rightleftarrows}} PMI^+ + I^- \quad \text{and} \quad I^- + I_2 \underset{k_{-2}}{\overset{k_2}{\rightleftarrows}} I_3^- \tag{S8}$$

The conditions of electro-neutrality and balances for the imidazolium cation and iodine atoms lead to the system of equations:

$$[PMII]_a = [PMII] + [PMI^+] \tag{S9}$$

$$[PMII]_a + 2[I_2]_a = [PMII] + 2[I_2] + [I^-] + 3[I_3^-] \tag{S10}$$

$$[PMI^+] = [I^-] + [I_3^-] \tag{S11}$$

$$K_1 = \frac{[PMI^+][I^-]}{[PMII]} = \frac{k_1}{k_{-1}} \tag{S12}$$

$$K_2 = \frac{[I_3^-]}{[I^-][I_2]} = \frac{k_2}{k_{-2}} \tag{S13}$$

The subscript "a" denotes the analytical concentrations of *PMII* and iodine. A numerical solution of this system of coupled equations can be obtained by restating it as a pseudo-kinetic problem, using the two coupled ordinary differential equations

$$\begin{cases} \dfrac{d[PMII]}{dt} = -k_1[PMII] + k_{-1}[PMI^+][I^-] \\ \dfrac{d[I^-]}{dt} = k_1[PMII] - k_{-1}[PMI^+][I^-] - k_2[I^-][I_2] + k_{-2}[I_3^-] \end{cases} \tag{S14}$$

Equation (S14), together with the balance equations (S9-S11) are integrated with a Runge-Kutta-Fehlberg algorithm using the initial conditions $[PMII](t=0) = [PMII]_a$ and $[I^-](t=0) = 0$. The time scale is determined by the arbitrary values of the rate constants, $k_1$ and $k_2$, but care must be taken to carry out the integration for a time span that is sufficiently long for establishing the equilibria.

The concentrations of the 5 species appearing in equation (S8) are calculated for trial values of $K_1$ and $K_2$, the values for $[PMII]_a$ and $[I_2]_a$ being the experimental values for the measurement of the dc conductivities (Figure 3B). The concentrations of the ions are then inserted into equations (S6) and (S7) to calculate the conductivity of each solution, using a third adjustable parameter, $f\delta^2$. A least squares minimization between the experimental and calculated values of the dc conductivity is carried out by varying the parameters $K_1$ and $f\delta^2$. The parameter $K_2$ is fixed at a high value, $1.0\times10^8$ M$^{-1}$ to keep the iodine concentration at a very low level, as neither UV/Vis nor Raman spectroscopy[14] have detected the presence of free iodine in the solutions.

A least squares minimum with parameter values $K_1$ = 5.2 M and $f\delta^2 = 7.2\cdot 10^{-19}\,\text{m}^2$ is obtained. Figure S3 shows the concentrations of the 5 species,

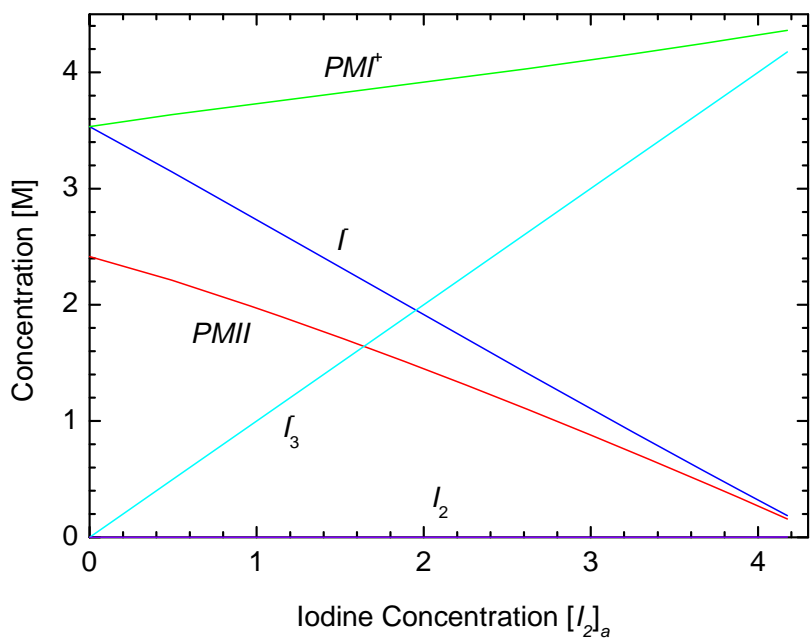

**Figure S3.** Concentrations of the ion-paired salt, *PMII*, of iodine $I_2$ and of the dissociated ions $PMI^+$, $I^-$, and $I_3^-$ calculated as a function of analytical iodine concentration $[I_2]_a$, using $K_1$ = 5.2 M and $K_2$ = $1.0\times10^8$ M$^{-1}$.

calculated as a function of $[I_2]_a$, using this value for $K_1$.

A distance $\delta = \sqrt{\delta_{I^-}^2 + \delta_{I_3^-}^2} = 9.3$ Å has been estimated from the bond length of triiodide. A factor $f = 0.83$ is calculated from the optimal value $f\delta^2 = 7.2 \cdot 10^{-19}\,\text{m}^2$. This implies that the rate constant $k_{ex}$ for the bond exchange reaction is close to the diffusion controlled limit. Alternatively, assuming $f = 1$ implies that $\delta$ is reduced to 8.5 Å. This smaller value of $\delta$ may be explained by the fact that not all collisions are head-on, but occur at an angle φ larger than zero (Figure S4).

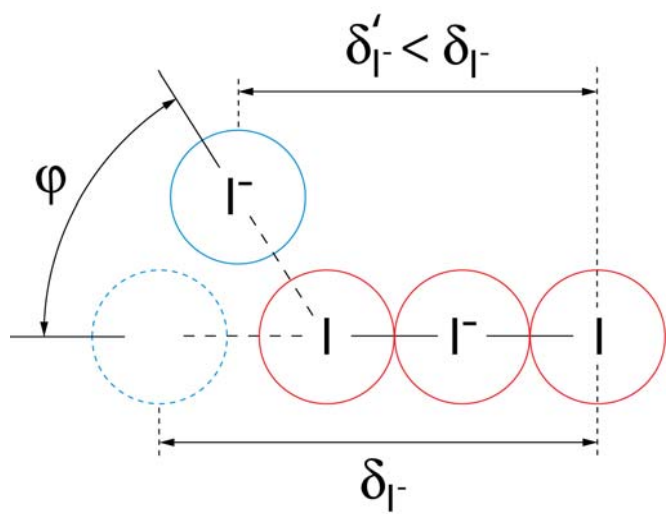

**Figure S4.** Illustration of the Grotthuss bond exchange reaction at the initial stage, occurring at a collision angle φ between the iodide and triiodide. $\delta_{I^-}$ is the maximum possible displacement of iodide. (The displacement of triiodide, $\delta_{I_3^-}$, is not illustrated for reasons of clarity).

## Section S3. Analysis of terahertz data

The decrease of the resonance amplitude at 2.03 THz resonance (Figure 3a) with increasing $[I_2]_a$, is correlated with the decrease of the concentration of the associated or ion-paired (IP) form of *PMII* and the increase of the dissociated (DIS) forms of *PMI$^+$/I$^-$ or PMI$^+$/I$_3^-$* (Figure S3). The measured complex dielectric function $\tilde{\varepsilon}(\upsilon)$ for the various concentrations $[I_2]_a$ of iodine is seen as resulting from the relative volume fractions and the dielectric functions $\tilde{\varepsilon}_{IP}(\upsilon)$ and $\tilde{\varepsilon}_{DIS}(\upsilon)$ of the two components IP and DIS of the mixture. (These dielectric functions do not depend on the iodine concentration and no distinction is made between the anions $I^-$ and $I_3^-$ as the imidazolium cation is assumed to be the main contributor to the signal[15]). The complex dielectric functions $\tilde{\varepsilon}(\upsilon)$, $\tilde{\varepsilon}_{IP}(\upsilon)$ and $\tilde{\varepsilon}_{DIS}(\upsilon)$ are related by a formula such as Bruggeman's effective-medium approximation[16,17],

$$v_{IP}\frac{\tilde{\varepsilon}_{IP}(\nu)-\tilde{\varepsilon}_{eff}(\nu)}{\tilde{\varepsilon}_{IP}(\nu)+2\tilde{\varepsilon}_{eff}(\nu)}+(1-v_{IP})\frac{\tilde{\varepsilon}_{DIS}(\nu)-\tilde{\varepsilon}_{eff}(\nu)}{\tilde{\varepsilon}_{DIS}(\nu)+2\tilde{\varepsilon}_{eff}(\nu)}=0. \qquad (S15)$$

Here $\tilde{\varepsilon}_{eff}(\upsilon)$ is the measured complex dielectric function of the mixtures of IP and DIS. Figure S5 shows the real (A) and imaginary (B) parts of the measured $\tilde{\varepsilon}(\upsilon)$ of the *PMII/I$_2$* mixtures up to 4.2 M (solid blue lines). Assuming $v_{IP}$ to be proportional to the IP concentration (Figure S3) we solve for values of $\tilde{\varepsilon}_{IP}(\nu)$ and $\tilde{\varepsilon}_{DIS}(\nu)$ simultaneously for all $[I_2]_a$ concentrations with a least squares fit, for each frequency. The calculated complex dielectric functions $\tilde{\varepsilon}_{IP}(\nu)$ and $\tilde{\varepsilon}_{DIS}(\nu)$ are shown in Figure S6, and $\tilde{\varepsilon}_{eff}(\nu)$ recalculated from equation (S15), using $\tilde{\varepsilon}_{IP}(\nu)$ and $\tilde{\varepsilon}_{DIS}(\nu)$ is shown in Figure S5 for each concentration (dotted red lines). The dielectric functions $\tilde{\varepsilon}_{IP}(\nu)$ and $\tilde{\varepsilon}_{DIS}(\nu)$ can each be modeled by two relaxation processes: relaxation of the orientational polarization induced by the electromagnetic radiation and resonance-induced polarization. The former is described by the Debye model which assumes an exponential relaxation of the polarization of the medium to equilibrium. For the latter we consider the motion of bound charges described by a classical damped harmonic oscillator. Plasma resonance relaxation due to freely moving electrons is not considered. The complex dielectric function then takes the form,

$$\tilde{\varepsilon}(\omega) = \varepsilon_\infty + \frac{\Delta\varepsilon}{1 - i\omega\tau} + \sum_{i=1}^{n} \frac{c_i}{\omega_{0,i}^2 - \omega^2 - i\gamma_i \omega}. \qquad (S16)$$

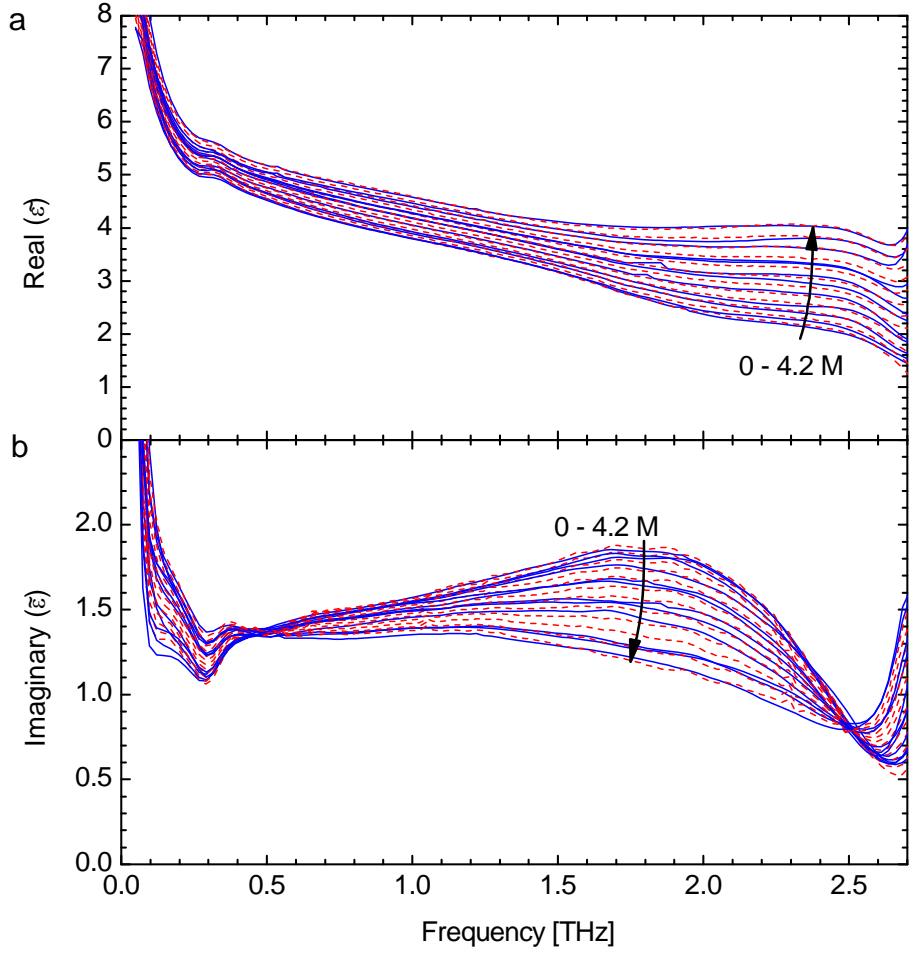

**Figure S5.** Real (A) and imaginary (B) part of the dielectric function of *PMII/I$_2$* mixtures at various iodine concentrations [$I_2$]$_a$ as a function of frequency. Solid blue lines: experimental data, $\tilde{\varepsilon}(\nu)$. Dotted red lines: $\tilde{\varepsilon}_{\text{eff}}(\nu)$ recalculated from the IP form $\tilde{\varepsilon}_{IP}(\nu)$ and the DIS form $\tilde{\varepsilon}_{DIS}(\nu)$ using equation (S15).

Here, $\Delta\varepsilon$ is the fraction of the dielectric amplitude loss due to Debye relaxation, $\varepsilon_\infty$ the dielectric constant at infinite frequency, $\omega$ the angular frequency, $\omega = 2\pi\upsilon$, and $\tau$ the Debye relaxation time. In the last term, $c$ is related to the oscillator strength, $\omega_o$ is the resonance angular frequency, $\gamma$ the damping constant, and $n$ the number of oscillators. A reasonable fit of equation (S16) to $\tilde{\varepsilon}_{IP}(v)$ and $\tilde{\varepsilon}_{DIS}(v)$, can be obtained with one Debye term and two, respectively three oscillator terms (Figure S6). The optimized values of the parameters are given in Figure S6. The Debye relaxation time of 1.21 ps (0.99 ps) for IP (DIS) is close to 1.9 ps and 2.6 ps obtained for the similar ionic liquid 1-ethyl-3-methylimidazolium triflate by Yamatomo et al.[15] and Asaki et al.[18] respectively. The fact that an additional oscillator term is needed to fit $\tilde{\varepsilon}_{DIS}(v)$, compared to $\tilde{\varepsilon}_{IP}(v)$, may be expected, as the shape of the imaginary part of $\tilde{\varepsilon}_{DIS}(v)$ is broader. This may be due to the looser nature of the dissociated forms *PMI$^+$/I$^-$ or PMI$^+$/I$_3^-$*, compared to the tighter bound ion-paired *PMII*.

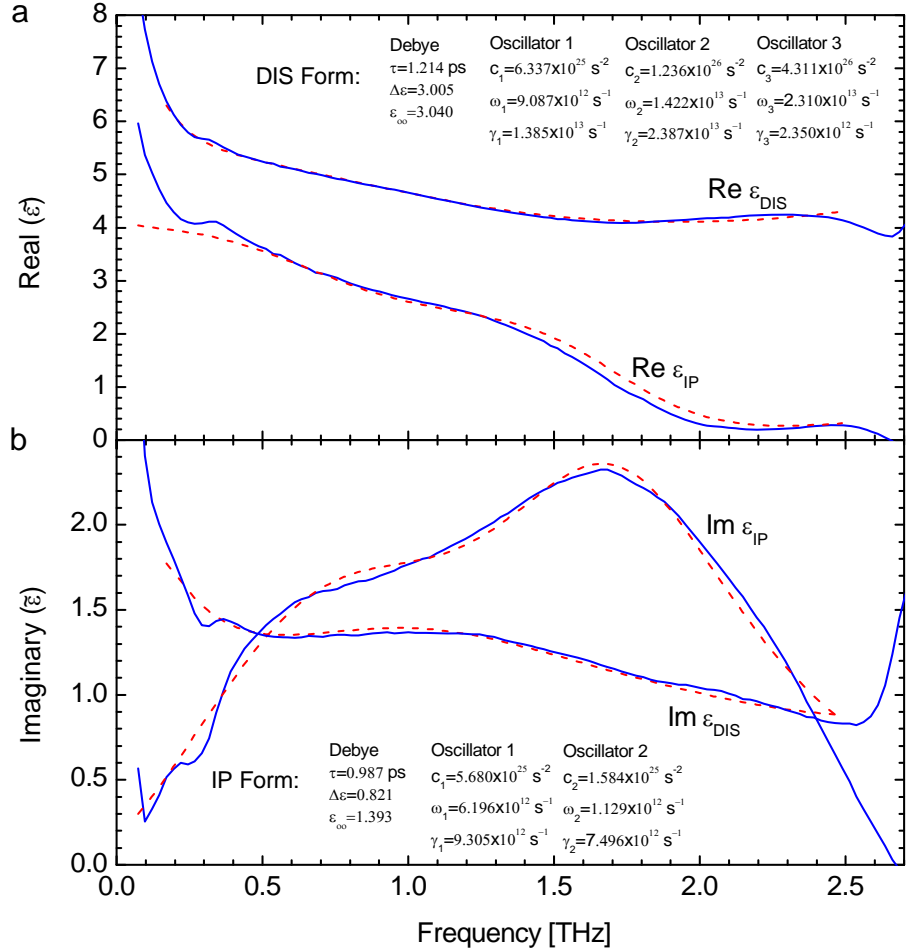

**Figure S6.** Real (a) and imaginary (b) part of the dielectric function of $\tilde{\varepsilon}_{IP}(\nu)$ (IP form) and $\tilde{\varepsilon}_{DIS}(\nu)$ (DIS form) as a function of frequency calculated from a least squares fit using equation (S15) (solid blue lines). Dashed red lines: Fits to $\tilde{\varepsilon}_{IP}(\nu)$ and $\tilde{\varepsilon}_{DIS}(\nu)$ using equation (S16). The IP form is fitted with 1 Debye term and 2 oscillator terms, and the DIS form is fitted with 1 Debye term and 3 oscillator terms. The optimized fitting parameters are indicated in the figure.

## Section S4. Supporting references